# Software-Defined Multi-Cloud Computing: A Vision, Architectural Elements, and Future Directions

Rajkumar Buyya and Jungmin Son

**Clou**d Computing and **Di**stributed **S**ystems (CLOUDS) Laboratory
School of Computer Science and Software Engineering
The University of Melbourne, Australia

**Abstract**

Cloud computing has been emerged in the last decade to enable utility-based computing resource management without purchasing hardware equipment. Cloud providers run multiple data centers in various locations to manage and provision the Cloud resources to their customers. More recently, the introduction of Software-Defined Networking (SDN) and Network Function Virtualization (NFV) opens more opportunities in Clouds which enables dynamic and autonomic configuration and provisioning of the resources in Cloud data centers.

This paper proposes architectural framework and principles for Programmable Network Clouds hosting SDNs and NFVs for geographically distributed Multi-Cloud computing environments. Cost and SLA-aware resource provisioning and scheduling that minimizes the operating cost without violating the negotiated SLAs are investigated and discussed in regards of techniques for autonomic and timely VNF composition, deployment and management across multiple Clouds. We also discuss open challenges and directions for creating auto-scaling solutions for performance optimization of VNFs using analytics and monitoring techniques, algorithms for SDN controller for scalable traffic and deployment management. The simulation platform and the proof-of-concept prototype are presented with initial evaluation results.

## 1. Introduction

Cloud computing is the fulfilment of the vision of Internet pioneer, Leonard Kleinrock [1], who said in 1969: *"Computer networks are still in their infancy, but as they grow up and become sophisticated, we will probably see the spread of 'computer utilities' which, like present electric and telephone utilities, will service individual homes and offices across the country."*

Cloud computing delivers infrastructure, platform, and software as a service,

which are made available to customers on demand as subscription-oriented services based on a pay-as-you-go model. This is supported by dynamic provisioning of computing resources of data centers using Virtual Machine (VM) technologies for consolidation and environment isolation purposes [2]. Many ICT infrastructure service providers including Amazon, Google, Microsoft, IBM, and Telstra are rapidly deploying data centers around the world to support customers worldwide. Furthermore, there is a wide array of services (e.g., standard and spot market priced VM/computing services in case of IaaS) offered by each provider for each model, and each service can be configured with different parameters.

Furthermore, the emerging IoT (Internet-of-Things) paradigm is enabling the creation of intelligent environments supporting applications such as smart cities, smart transportations, and self-driving vehicles. They demand services that are delivered with low latency to meet their QoS (Quality-of-Service) requirements. This triggers the need for distributed Clouds, which are placed near edge-oriented IoT applications.

### 1.1 The Programmable Network Cloud: Challenges and Requirements

Offering a variety of services creates a challenge for service providers to enforce committed Service Level Agreements (SLAs), which are made up of QoS expectations of users, rewards, and penalties in case it is not achieved. Since SLA establishment is legally required for compliance and has potential impact on revenue, meeting these SLAs is a primary concern of service providers. One elementary approach to guarantee SLA compliance is provisioning of resources for peak requirements of a given service. This approach is not economically viable, i.e., over-provisioning increases the cost of running the service as these provisioned resources are often underutilized.

As data centers are made up of large numbers of servers and switches, they consume significant amounts of energy. According to the Natural Resources Defense Council, **data centers in the U.S. consumed** about 90 billion kilowatt-hours of electricity in 2013, which is roughly **twice of the electricity consumption** in New York City. The energy cost incurred by data centers in 2014 was estimated at **54 billion dollars**. It is reported that energy expense typically accounts for more than **75% of Cloud data center operating expenditure**.

Operators can utilize a new paradigm called "Programmable Network Cloud" [3], which provides scalable services efficiently through a combination of Software-Defined Networking (SDN) [4] and Network Functions Virtualization (NFV) [5] along with integrated use of virtualized computing and storage resources. This enables efficient sharing of resources and helps in meeting SLA requirements while minimizing the operational cost.

SDN allows separation of the control and configuration of the network routes from the forwarding plane (provided by networking devices) [6]. This offers flexibility to the network control plane by enabling the network to be easily adapted to changes via the software called *controller*. OpenFlow [7] is a de facto standard interface for SDN controllers. It allows controllers to communicate with the forwarding plane and makes dynamic changes to the network. This real-time responsiveness to traffic demand is an effective feature to deal with the dynamic nature of the telecom operators' networks, as huge numbers of network resources are constantly joining and leaving the network.

NFV concerns the migration of network functions, such as load balancing, network address translation, and firewalls, to the software layer to enable better interoperability of equipment and advanced network functions. While SDNs focus mostly on separating the control and forwarding planes, NFV focuses on other types of networking functions that are commonly embedded on networking devices or realized in the form of *middleboxes*, i.e., hardware appliances that implement a specific network function such as filtering unwanted or malicious traffic. The deployable elements of NFV are known as *Virtualized Network Functions (VNFs)* that are hosted in containers/virtual machines in Clouds and benefit from its elasticity.

Nevertheless, offering a variety of services in a timely manner while maximizing user experience across the world and at the same time minimizing operational costs (OpEx) creates a number of challenges in the design and realization of the Programmable Network Cloud. New services often need resources and application components that reside in different Clouds or network domains, and in this aspect, it diverges significantly from the behavior of traditional services deployed in traditional siloed data centers. The current centralized data centers impose a number of limitations on the way network and application services are delivered. The utilization of multiple distributed Clouds brings applications closer to the access network/devices, which will enable substantially lower latencies [3]. This is not a trivial task as multiple data centers with different platforms need to be managed simultaneously, with efficient approaches for resources provisioning, service placement, and distribution of the workload.

SDN and NFV can serve as building blocks for achieving traffic consolidation thus saving cost while also supporting QoS at network level in Cloud data centers [3]. However, this requires joint host-network consolidation optimization techniques that not only minimize the energy cost by consolidating network traffic and VMs onto the least number of links and hosts, but also avoid negative impacts on user experience.

In addition, the ability to allocate resources elastically is one of the major advantages of a Programmable Network Cloud that can significantly reduce OpEx. Although by leveraging the Cloud, network operators can scale resources up or down according to changes in demand, the key question is when auto-scaling should be triggered and how much capacity shall be added or removed. With millions of nodes operating in the network, manual management of elasticity is not feasible. Hence, there should be an automated approach in place that utilizes Big Data analytics along with efficient monitoring to detect/predict anomalies in service performance and act accordingly.

Moreover, provisioning of new resources and services taking months or even weeks to be performed is not acceptable in the current competitive markets. Therefore, a Cloud agnostic deployment configuration management (DevOps) solution that automates the deployment of VNFs and applications across multiple data centers running different platforms is of extreme importance.

Without SDNs, dynamic and quick reaction in response to variation in application condition while maintaining service quality cannot be achieved at the network level. It continues to be challenging, as existing SDN controllers are still rudimentary and they need scalable and automated algorithms for achieving integration of such features to enable performance-driven management. SDN and NFV technologies do not, by themselves, solve any of the problems related to QoS nor represent an improvement over traditional networks if the controller is not properly instrumented. Therefore, new algorithms and techniques need to be developed and incorporated in the controller to realize the benefits of the Programmable Network Cloud.

The main objective of this paper is to explore *the above-mentioned challenges by going beyond traditional Cloud services and proposing solutions* for auto-scaling, cross-cloud cost and SLA-aware VNF/VM provisioning and consolidation, scalable SDN-controller traffic management, Cloud agnostic deployment configuration management techniques, use of open source initiatives, and Data Analytics-enabled VNF performance tuning in multi-cloud computing environments.

**1.2 Research Methodology**

A methodology for solving the problem of cost-efficient and SLA-aware management of geographically distributed Cloud data centers where network functions and applications can be instantly deployed and managed to meet acceptable service levels of customers world-wide is as follows:

- Define an architectural framework and principles for cost-efficient and SLA-aware management of future network resources and applications distributed across multiple Cloud data centers.
- Propose new algorithms for cost-efficient and SLA-aware management of resources by joint host and network optimization.
- Create new data analytics techniques for auto-scaling, performance tuning, and failure recovering of network services.
- Propose techniques for autonomic VNF (Virtualized Network Function) and application deployment and their management across multiple Clouds to accelerate new service deployment processes.
- Propose new algorithms for autonomic VNF composition and management aiming at enabling efficient execution of applications by increasing the efficiency in the utilization of the data center network. This includes techniques for placement and consolidation of VNF compositions.
- Develop new algorithms and optimization techniques for SDN controllers for traffic and deployment management to decrease cost of data centers while honoring Service-Level Agreements.
- Design and develop a proof-of-concept software platform incorporating the above techniques and demonstrate their effectiveness through a series of applications.

The rest of this paper is organized as follows: First, a brief survey on the existing literature in SDN usage in Clouds is presented. Next, we propose a system ar-chitecture of Software-Defined Multi-Cloud platform and explain its elements in detail, followed by the simulation framework implementation and some preliminary experiments and results to show the potential effectiveness of the proposed approach. Finally, the paper concludes with discussion on future directions.

## 2. State-of-the-art in SDN usage in Clouds

Although recent research has addressed the application of SDN concepts in Clouds [9][10][29][33], little attention has been given on how to enable network performance allocation to different customers with diverse priorities beyond best effort. Therefore, the user segmentation and bandwidth allocation proposed in this paper make novel contributions for the topic of Cloud networking.

On the topic of energy efficiency, existing works [11][12][13] focus mainly on improving energy efficiency of servers. Only a few recent works started to look at the problem of energy efficient Cloud networking [8]. As the available works are in early stage, they make too many assumptions that do not hold in real data centers.

Regarding NFV, previous research [14][15][16][17] investigated the problem of middlebox virtualization. Some works proposed migration of middleboxes functionalities to the Cloud, whereas others focused on enabling virtual middleboxes by using SDNs. These approaches do not target energy efficiency of the infrastructure, but rather best effort while moving the functionalities from hardware appliances to Cloud virtual machines.

Virtualization technology enables energy-efficiency via workload or computing server consolidation [11][13]. The next research step in this direction concerns algorithms that jointly manage consolidation of VMs and network traffics to the minimum subset of hosts and network resources.

Furthermore, many energy-efficient resource allocation solutions proposed for various computing systems [18][19][20] cannot be implemented for Software-Defined Clouds. This is because they do not explicitly consider the impact of their solution on SLA requirements of different classes of users with diverse priorities. Hence, they do not emphasize SLA and energy-aware resource management mechanisms and policies exploiting VM and network resource allocation for Cloud services offered in different QoS.

## 3. System Architecture

Figure 1 shows the high-level unified architecture that accelerates and simplifies the deployment of network functions along with other applications in distributed Cloud data centers. The architecture leverages the state-of-the-art technologies and paradigms to deliver reliable and scalable network functions while minimizing the operational cost of data centers and meeting QoS requirements of users. In the fol-lowing sections, we present components of the framework, their related research problems, and potential solutions.

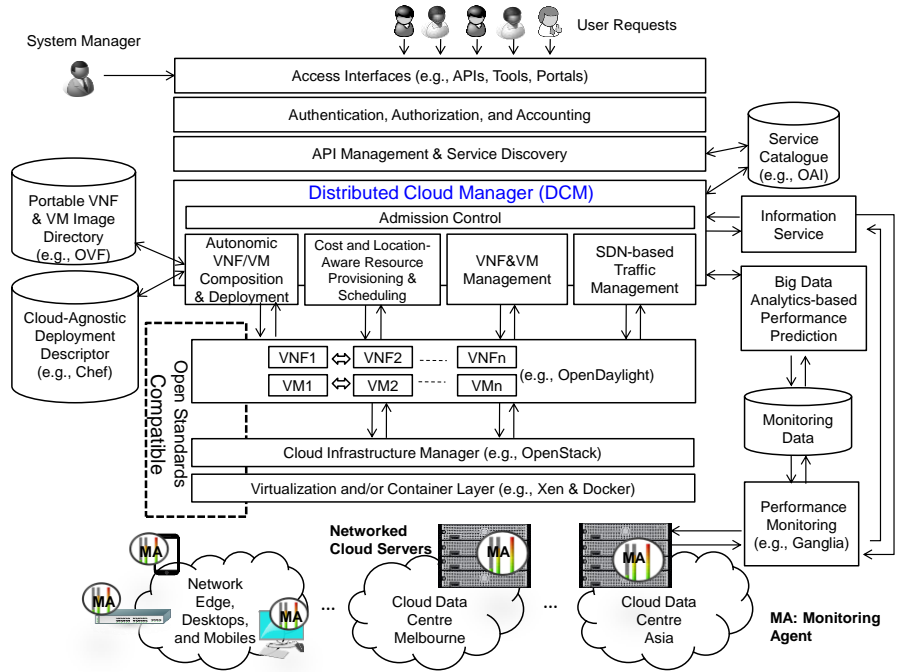

Figure 1: System Architecture for Distributed Cloud Data Centers with NFV.

The framework is designed in such a way that it enables network service placement across multiple platforms in distributed Multi-Cloud environments. It allows deployment and migration of VNFs and user applications across Cloud platforms. This is best achieved by the following principles:

- Use of open Cloud standards and execution environments such as OpenDaylight, OpenFlow, OpenStack, Open Virtualization Format (OVF), and Open Container Format (OCF).
- Use of abstractions and virtualization techniques that allow seamless deployment and migration of services across multiple platforms.
- Utilization of Cloud characteristics of elasticity and auto-scaling to adapt the platform to changes in demand.
- Use of automation techniques and tools including DevOps to accelerate new service deployments.
- Use of open APIs formats for service exposures and discovery to enable 3rd party innovations.

As noted in Figure 1, the input of the system consists of *static* and *dynamic requests*. *Static requests* are issued by the System Manager and consist of instructions for provisioning and configuration of VMs/containers and VNFs. Once

VMs/containers that host applications and VNFs that host network operations are placed and the network configuration is completed, the system is ready to receive *dynamic requests* that are issued by end users.

The **Distributed Cloud Manager (DCM)** is the core component of the architecture and consists of a set of subcomponents with specialized functions. The *Cost and Location-aware Resource Provisioning and Scheduling* component uses monitoring data as input and aims at improving efficiency of the data center while minimizing SLA violations via efficient VM/container and traffic placement and consolidation across multiple Clouds. The architecture further minimizes SLA violations and (energy) cost by dynamically finding the shortest path for each network flow via the *SDN-based Traffic Management* component. *Monitoring Data* is collected regarding the servers' utilization and service performance. Such information is obtained by *Monitoring Agents (MA)* stored in a database that is available to DCM components through the *Information Service* component. To ensure the highest level of SLA satisfaction, DCM utilizes the *Autonomic VNF&VM Management* component that is responsible for auto-scaling and failure recovery. Timely scaling and recovery is made possible with the help of the *Performance Prediction* component that consumes historical data (collected using benchmarking of VNFs deployed on various VM types) and predicts performance anomalies. Finally, the *Autonomic Composition, Configuration and Deployment Management* component is responsible to store the deployment information and simplify the deployment of VNFs using preconfigured images.

## 4. Architectural Elements, Challenges, and Directions

In this section, we discuss various key elements identified in system architecture along with challenges and issues involved in their realization.

### 4.1 API Management and Service Discovery

A key issue with the way services provided over the Web are consumed is the method of access. No assumption can be made these days about the nature of access devices or services that utilize them. Indeed, one cannot even assume that a service is being consumed by humans as Web services can be made up of many disparate simpler services. Such microservices promote reuse and isolation, resulting in higher tolerance to failures.

The *Service Catalogue* component serves as a directory/registry maintaining information related to the available services and their access methods. It can be accessed by users and other services/brokers to discover them. The *API Management and Service Discovery* component uses information from the Catalogue to determine which particular services are required for each incoming request, and to keep track of the health of the corresponding microservices. New algorithms can be developed to enable composition of the complex services and workflow execution required to perform the activities of other components of the architecture.

## 4.2 Admission Control

Given that Cloud infrastructure is composed of a finite amount of physical resources, it is unrealistic to assume that there is an infinite amount of computing and network power available for processing an arbitrary number of requests. Thus, if the number of incoming requests exceeds the infrastructure capacity, requests will experience delays and poor QoS, resulting in violation of SLAs. Thus, the key research challenge to prevent QoS degradation is *SLA-Aware Admission Control* algorithms.

Because providers are subject to penalties for SLA violations, it is necessary to evaluate which requests should be accepted and which ones should be rejected. To enable a rational decision and **to increase provider revenue, the decision should be based on user segmentation**. In this case, requests belong to users with different service levels, and requests from higher levels have priority over the lower level ones for admission purposes. Regardless the service level, it is important to avoid starvation, a problem where requests from lower levels are never accepted, as this leads to abandonment of the service by customers.

## 4.3 Cost and Location-Aware Resource Provisioning and Scheduling

The existing centralized Cloud deployments limit the way certain network services can be delivered. In order to offer satisfactory Quality of Experience (QoE), providers require resources that reside in Cloud data centers distributed across different geographical locations. The load on each individual Cloud can vary with time, sometimes in unexpected ways. This necessitates a solution that manages multiple resource pools with different platforms that can exploit elasticity and auto-scaling properties of Clouds and at the same time is equipped with optimization algorithms that both maximize the QoE of users and minimize the operational cost.

Over-provisioning of resources (e.g., CPU and memory) is one of the major causes of cost escalation although the average utilization of servers is reported to be between 10-30% for large data centers [21]. Hence, to improve the server cost efficiency, VM/container placement, consolidation, and migration techniques play an important role. Likewise, provisioning of network capacity for the peak demand leads to bandwidth wastage. With the emergence of SDN, it is possible to investigate approaches to dynamically manage bandwidth per flow and thus consolidate network traffic. However, poor traffic and VM/container consolidation can have a negative effect on the quality of service (QoS) that clients receive from the system and it can impact the system profitability. Therefore, optimization algorithms are required that can minimize cost via consolidation while still meeting QoS requirements.

The Host-Network resource provisioning problem is a variant of the multi-commodity flow problem and can be solved by linear programming solvers. As they have high computational complexity, online greedy bin-packing algorithms or heuristics can be used to substitute the optimal solution. They not only find the minimum subset of hosts and network resources that can support a given workload, but also dynamically identify resources that satisfy SLA requirements of users such as latency by considering users locations. For efficient overbooking and VM/container and flow consolidation, workload correlation analysis and predic-tion methods can be utilized [22].

To address the above issues, *cost-efficient joint VM/container and VNF provisioning* needs to be investigated in the context of this framework component. Furthermore, the problem of *resource and workload placement* needs to be addressed in the context of SDN in Clouds, as it has not been explored so far in the literature, resulting in a lack of solutions that can handle the scale of distributed Cloud infrastructure while meeting SLA demands of applications. **Therefore, it is worth to investigate optimization techniques that simultaneously optimize the VM/Container placement and traffic consolidation to jointly maximize SLA satisfaction and minimize the cost (including energy)**.

Furthermore, client requests directed to the applications deployed on the platform need to be scheduled in a timely manner so that the expected QoS of the application is achieved. The application may contain several VNFs that are subject to different load levels. Thus, it is important that the request is scheduled to a VNF (or VNF path) that can complete its execution within the agreed service level. Considering the importance of the scheduler component for the success of the framework, it is crucial to consider the following three key elements:

The first element is an *analytical model* for SDN-enabled Cloud data centers that combines *network-aware scheduling* and *load balancing*. It is capable of capturing SDN principles and at the same time sufficiently flexible to handle priorities defined for each flow. The analytical modelling of SDN has not been investigated deeply and it is attempted only in a few research works in the literature [21][23]. These proposed models generally cannot be extended to more than one switch in the data plane and have not considered the priority of flows. Therefore, it is necessary to build a model based on priority networks [23] that can deal with the aforementioned shortcomings and can efficiently be used for analysis of SDN networks and validation of results from experiments conducted via simulation.

The second element, *network-aware scheduling and load balancing* in distributed Clouds, serves users with different priorities. The main objective is the exploration of SDN capabilities for dynamic bandwidth management on a per flow basis for each class of users. SDN can prioritize traffic of users with higher classes by dynamically configuring queues in switches, which can be utilized by resource allocation policies. Combining this problem with the problem of capacity management including scaling, VM/VNF placement and migration makes this research topic even more challenging. Finally, in order to move towards more realistic scenarios, the original problem invested above should be extended to simultaneously consider the two objectives of **adhering to SLA constraints of different classes of users and minimizing operational cost in the data center**. This can be achieved with the utilization of the network topology and the relative distance between hosts during the formulation of the scheduling problem.

**4.4 Autonomic Configuration, Deployment Management, and VNF Composition**

A high-level network service is usually composed of many smaller services that are well described and widely used in computer networks. Implementation of traffic policies normally requires the combination of services such as network address translation (NAT), firewalling, load balancing, and LAN and WAN optimization. Each of these services manipulates the input network packets and modifies it to achieve its goals. These operations are usually CPU-intensive and occur for hundreds or thousands of packets per second. In current competitive markets, service rollout that takes months or even weeks is not acceptable. Hence, it is important that the service composition can be performed efficiently and automatically. It can be realized by defining VNF composition and its role for efficient execution of applications. A VNF composition is called a VNF Graph and enables advanced networking services. The automatic composition methods can adopt AI planning techniques. Also, online optimization algorithms are necessary for placement and consolidation of VNF Graphs and for scheduling and load balancing of user requests on VNF Graphs across multiple Cloud data centers. Each virtualized function may be scaled out independently to support the observed demand. **Thus, auto-scaling and elasticity capabilities can be achieved.**

Once the configuration of VNF graphs is finalized, and the decision on mapping of VM/container instances to hosts has been made, the deployment and configuration metadata has to be persisted in such a format that can later be used to accelerate the redeployment of VMs/containers and VNFs in case of a failure in a new platform. Storage of metadata also enhances reusability, as it enables functionalities similar to those provided by automation tools such as Chef [24], which utilizes reusable definitions and data bags to automate deployment. Therefore, **a new approach is necessary to persist specifications of required VNFs and their configuration information**, such as scaling settings, in a format we termed as Deployment Descriptor by leveraging and extending current DevOps technologies. The format has to include instance description (e.g. name, ID, IP, status), image information, etc. This metadata is used by the Configuration Manager and DCM to manage the whole stack of VNFs even if they are deployed across multiple Clouds. One possible solution is to extend existing vendor-agnostic formats such as the Open Virtualization Format (OVF) [25], which is an open specification for the packaging and distribution of virtual machine images composed of one or more VMs. OVF can be further enhanced to facilitate the automated and secure management of not only virtual machines in our system but also the VNF as a functional unit.

**4.5 Performance Management of VMs and VNFs**

Apart from physical resources management, virtual resources also require monitoring. VMs executing user applications and VNFs providing networking support for applications are CPU-intensive and are subject to performance variation caused by a number of factors, such as overloaded underlying hosts, failure in hardware components, or bugs in their underlying operating systems. Therefore, to keep a certain level of QoS, VMs and VNFs need to be managed such that their status is constantly monitored, and when problems are detected, corrective actions need to be taken automatically to avoid malfunctions. Considering the importance of autonomic management, **auto-scaling mechanisms** can play important role for VNF Graphs. The auto-scaling mechanism consists of two components:

- **Monitoring** component that keeps track of performance of the VNF Graph.

Data collected by the monitoring is used to enable performance prediction.

- **Profiling** component that collects information to build correlations regarding the VNF computing requirements and service time. The main challenge in this regard is the lack of benchmarking tools developed specifically for VNFs. New benchmarks can be proposed to measure performance criteria such as VNFs' throughput, time to deploy, and migration impacts on SLAs. The benchmark can help to design auto-scaling policies that maintain a certain level of QoS once any anomaly in the expected performance is detected.

**4.6 Big Data Analytics-Based Performance Prediction for Capacity Planning and Auto-scaling**

Several functionalities of the proposed architecture rely on data analytics for proper operation. However, the data generated by the infrastructure and consumed by the proposed architecture has characteristics of volume (because there are thousands of components—from infrastructure to application level—that need to be monitored), velocity (data is being continuously generated, and needs to be processed quickly to enable timely action to incidents), and variety (data is generated in various formats). These are the characteristics of **Big Data analytics**, which need to be handled with specialized systems and algorithms.

The collected data can be used for many purposes inside the architecture. Firstly, it can be used in new methods based on Big Data to detect performance anomalies in VMs, VNFs, and in the network as a whole, and to identify the root cause of the anomaly. This can be done after defining the set of required QoS criteria and metrics; and identifying and modelling the dependency between those criteria. This helps in improving the accuracy of the monitoring system by locating the root cause of a failure. The activity also looks into developing recovery plans related to performance issues caused by hosts and link failures. Big Data analytics frameworks such as Spark can be exploited for rapid detection of anomalies and identify root causes that assist in data center's capacity planning and auto-scaling.

**4.7 Software-Defined Network (SDN)-based Traffic Management**

SDN is a prominent technology for data center traffic management that provides flexibility to dynamically assign a path (network capacities) to a data flow through APIs. However, knowing that the operators' networks are dynamic, computing and updating the shortest path per flow must be performed in efficient way. One possible solution is the design and optimization of vertex-oriented graph processing techniques to dynamically find the shortest paths. For this purpose, solutions can be implemented in Apache Giraph [26]. The dynamism and changes in the data center network can be evaluated under different topologies, and graph processing can be applied to recalculate the shortest paths in response to changes. A scalable graph partitioning and processing algorithm is also necessary to takes advantage of Cloud's elasticity. It helps in building efficient controllers that properly utilize elasticity of Cloud computing and provide forwarding rules in a timely manner. In detail, the issue can be investigated in twofold:

- **Efficient heuristics for graph partitioning** that support a vertex-oriented execution model [27] and suit data center network routing problems, which enables scalable and efficient computations.

- **Algorithms for detection of similarities** in network conditions and configurations including topology, traffic, and priority settings (the current and previous network setting retrieved from the history) to accelerate the process of making routing decisions for current network settings.

These techniques can be incorporated into OpenFlow/OpenDaylight stacks for efficient management of network traffic.

### 4.8 Authentication, Authorization, Accounting, and System Security

Achieving security features such as confidentiality (protecting data from one user from others), availability (make the application available to legitimate by blocking malicious users), and protection against **Distributed Denial of Service (DDoS)** attacks is crucial for any solution deployed in the Internet, and they are specialty important when the solution can be distributed along multiple Clouds. Authorization protocols such as OAuth2 [28] can be used to guarantee that VNFs and VMs are seamlessly deployed across multiple Clouds if necessary.

Besides single sign in capabilities, this module also performs DDoS detection and mitigation. This is a critical feature, because in a dynamic resource provisioning scenario, increase in the number of users causes an automatic increase in resources allocated to the application. If a coordinated attack is launched against the Cloud provider, the sudden increase in traffic might be wrongly assumed to be legitimate requests and resources would be scaled up to handle them. This would result in an increase in the cost of running the application (because the provider will be charged for these extra resources) as well as waste in resources.

In this module, **automatic identification of threats** can be implemented by checking the legitimacy of requests. For this, the module needs to perform analytics on data center logs and feedback from intrusion detection systems, which leads to prevent the scaling-up of resources to respond to requests created with the intention of causing a Denial of Service or other forms of cyber-attacks. The module is also able to distinguish between authorized access and attacks, and in case of suspicion of attack, it can either decide to drop the request or avoid excessive provision of resources to it. To achieve this, techniques already in use for detection of DDoS attacks can be adapted to be able to handle the unique characteristics of Cloud systems.

## 5. Performance Evaluation - Platforms and Sample Results

In this section, we present both the simulation and real software platforms that support a programmable network Cloud for efficient network services via SDN and NFV and show some of our preliminary experiment results.

### 5.1. Simulation Platform for SDN and NFV in Clouds

In order to simulate NFV and SDN functionalities in Clouds, we have implemented additional modules in CloudSimSDN simulation toolkit [30]. CloudSimSDN has been developed based on CloudSim [31] to support testing SDN functionali-ties, such as dynamic flow scheduling, bandwidth allocation, and network programmability for Cloud data centers. On top of CloudSimSDN, our extension modules for NFV support creating VNFs in a data center and composition of multiple VNFs to form a chain of network functions. It also supports altering forwarding rules in SDN switches to enforce specific packets to transfer through VNF chains using defined policies.

As the simulation tool is developed upon CloudSimSDN, it is capable of all the features and functions provided by CloudSimSDN and CloudSim. We can measure the response time of each workload compositing of computing and networking jobs and the energy consumption for each host/switch and the entire data center. It also supports to implement different policies and algorithms for VNF placement and migration which decides a physical host to place a VNF.

**5.2. System Prototype**

We also developed a software platform named SDCon[1] for empirical evaluation of the proposed architecture in a small-scale testbed [32]. The prototype is implemented upon various open-source software such as OpenStack, OpenDaylight, and OpenVSwitch. The software integrates Cloud manageability and network provisioning into a single platform, which can be exploited for joint host-network optimization, network bandwidth allocation, dynamic traffic engineering, and VM and network placement and migration.

The system prototype is implemented on the in-lab testbed consisting of 9 servers and 10 software switches. We utilize Raspberry-Pi low-cost embedded computers for the software switches, by plugging USB-Ethernet adapters and running OpenVSwitch software. The testbed with the deployed SDCon can easily perform the empirical experiments for SDN usage in the Cloud context.

**5.3 Evaluating VNF Auto-Scaling Policy in Simulation**

In previous subsection, we present the simulation framework for VNF and SDN functionalities in Clouds. Upon the implemented simulation, we undertook preliminary experiments to show the validity of the presented vision in this paper. The experiment is designed to see the effectiveness of VNF auto scaling policies in the VM network performance and energy consumption.

We created 34 VMs and 4 VNFs in a data center with 128 hosts connected with 8-pod fat-tree network topology. All data transmissions between VMs are enforced to pass through a chain of 4 VNFs. We synthetically generate workloads for 10 minutes including compute (CPU) workload in VMs and network transmission to another VM. We measure CPU processing time, network processing time, and overall response time for each workload in two cases. For the first case, we enabled auto-scaling algorithm for VNFs, which simply adds more VNFs in the data center once the current utilization of a VNF reaches the threshold (70%). In the second case, the auto-scaling algorithm is disabled, so that the number of VNFs are never changed.

---

[1] Source code available at: https://github.com/Cloudslab/sdcon

Table 1 presents the total capacity of all VNFs at the end of the experiment. With the auto-scaling policy enabled, it keeps adding VMs utilized for VNFs throughout the experiment once the utilization of a VNF reaches the threshold. In the end, the auto-scaling policy added 5 more VMs in total for overloaded VNFs, whereas the number of VMs has not changed when we disabled auto-scaling policy.

Table 1: VNF capacity with different auto-scaling policies

|  | Initial | Auto-Scaling enabled | Auto-Scaling disabled |
|---|---|---|---|
| Number of VMs used for all VNFs | 4 | 9 | 4 |

Table 2 present CPU, network, and overall processing time of 90,000 workloads sent for 10 minutes. As shown in the table, the auto-scaling policy for VNFs reduced the average network transmission time by 22.4% from 1579 msec to 1226 msec, which leads to the improvement of the overall response time by 22.2%. As the number of VMs for VNFs has been increased by auto-scaling policy, the packets processed faster with higher capacity of VNFs with the auto-scaling enabled. Although this experiment is preliminary to show the effectiveness of the simulation framework, the toolkit can be used for further NFV and SDN studies in Clouds context to evaluate various composition and migration algorithms, traffic management, and joint compute-network optimization within a single data center, or for Multi-Clouds.

Table 2: Average CPU processing, network transmission, and response time of workloads

|  | VNF Auto-Scaling enabled | VNF Auto-Scaling disabled |
|---|---|---|
| CPU Processing | 27 msec | 27 msec |
| Network Transmission | 1226 msec | 1579 msec |
| Response Time | 5018 msec | 6454 msec |

## 6. Summary and Conclusions

In this paper, we presented a vision, a model architecture, elements, and some preliminary experimental results of Software-Defined Multi-Cloud Computing. Many opportunities and open challenges exist in the context of NFV and SDN in Cloud computing. As we presented in this paper, joint resource provisioning and scheduling in Software-Defined Cloud computing which considers VM, VNF, storage, and network performance and capacity are one of the most challenging topic. The integration of such resources and modules with a holistic view [32] is necessary in the future Software-Defined data center architecture, so that all the resources can be managed in autonomically. Also, traffic engineering with scalable NFV and programmable SDN is worth to investigate in order to optimize the network performance, SLA fulfillment, and the operational cost of a data center including electricity consumption.

The evaluation of the new architecture and its elements along with proposed or new approaches addressing their challenges can be fostered using our simulation (CloudSimSDN) and/or empirical platforms (SDCon). For a large-scale evaluation, our simulation toolkit can speed up the evaluation process with various measurements. For a practical proof-of-concept experiment, the empirical evaluation platform can be exploited to see the effectiveness of the proposed methodology in the real world. More details of the simulation and empirical platforms have been discussed in [32].

**Acknowledgements:** We acknowledge Dr. Rodrigo Calheiros, Dr. Amir Vahid Dastjerdi, and Dr. Adel Nadjaran Toosi for their contributions towards various ideas presented in this paper. This work is partially supported by an Australian Research Council (ARC) funded Discovery Project.

# References

[1] L. Kleinrock. A Vision for the Internet. *ST Journal of Research*, 2(1):4-5, Nov. 2005.
[2] P. Barham et. al. Xen and the Art of Virtualization. In *Proc. of 19th ACM Symposium on Operating Systems Principles (SOSP 2003)*, Bolton Landing, USA, 2003.
[3] The programmable network cloud. Ericsson White paper. Uen 288 23-3211 Rev B | December 2015, URL: http://www.ericsson.com/res/docs/whitepapers/wp-the-programmable-network-cloud.pdf
[4] Open Network Foundation. *Software-Defined Networking: The New Norm for Networks.* White Paper, April 2012.
[5] M. Chiosi et al. *Network Functions Virtualisation—Introductory White Paper*. White Paper, October 2012. URL: http://portal.etsi.org/NFV/NFV_White_Paper.pdf
[6] L. Barroso and U. Holzle. *The datacenter as a computer: An introduction to the design of warehouse-scale machines.* Synthesis lectures on computer architecture, 4(1):1–108. Morgan & Claypool, 2009.
[7] M. Jarschel et al. *Modeling and Performance Evaluation of an OpenFlow Architecture*, Proc. of the 23rd International Teletraffic Congress, 2011.
[8] K. Zheng, X. Wang, L. Li, X. Wang. *Joint Power Optimization of Data Center Network and Servers with Correlation Analysis.* Proc. of the 33rd IEEE Intl. Conf. on Computer Communications (INFOCOM'14), Toronto, Canada, April 2014.
[9] T. Koponen et al. *Onix: A distributed control platform for large-scale production networks.* Proc. of the 9th USENIX conference on Operating systems design and implementation (OSDI), Broomfield, USA, October 2010.
[10] C. Monsanto, J. Reich, N. Foster, J. Rexford, and D. Walker. *Composing Software-Defined Networks.* Proc. of the 10th USENIX Conference on Networked Systems Design and Implementation (NSDI), Lombard, USA, April 2013.
[11] A. Beloglazov, R. Buyya, Y. Lee, and A. Zomaya. *A Taxonomy and Survey of Energy-Efficient Data Centers and Cloud Computing Systems*. Advances in Computers, 82:47-111, Elsevier, March 2011.

[12] B. Guenter, N. Jain and C. Williams. *Managing cost, performance, and reliability tradeoffs for energy-aware server provisioning*, Proc. of the IEEE Conference on Computer Communications (INFOCOM), Shanghai, China, 2011.

[13] A. Beloglazov and R. Buyya. *Managing Overloaded Hosts for Dynamic Consolidation of Virtual Machines in Cloud Data Centers Under Quality of Service Constraints*. IEEE Transactions on Parallel and Distributed Systems, 24(7):1366-1379, 2013.

[14] J. Sherry, S. Hasan, C. Scott, A. Krishnamurthy, S. Ratnasamy, V. Sekar. *Making middleboxes someone else's problem: network processing as a Cloud service.* SIGCOMM Computer Communication Review, 42(4):13–24. ACM, October 2012.

[15] Z. Qazi, C.-C. Tu, L. Chiang, R. Miao, V. Sekar, M. Yu, *SIMPLE-fying Middlebox Policy Enforcement using SDN*. Proc. of the ACM 2013 Conference on SIGCOMM, Hong Kong, China, August 2013.

[16] A. Gember et al. *Stratos: A network-aware orchestration layer for virtual middleboxes in Clouds*. URL: http://arxiv.org/abs/1305.0209, 2014.

[17] J. Hwang, K. Ramakrishnan, and T. Wood. *NetVM: High performance and flexible networking using virtualization on commodity platforms.* Proc. of the 11th USENIX Symp. on Networked Systems Design & Implementation, USA, April 2014.

[18] H. Aydin, R. G. Melhem, D. Mossé, and P. Mejía-Alvarez. *Power-Aware Scheduling for Periodic Real-Time Task*. IEEE Transactions on Computers, 53(5): 584-600, May 2004.

[19] J. S. Chase, D. C. Anderson, P. N. Thakar, and A. M. Vahdat. *Managing Energy and Server Resources in Hosting Centres*. Proc. of the 18th ACM Symposium on Operating System Principles, Banff, Canada.

[20] H. Zeng et. al. *ECOSystem: Managing Energy as a First Class Operating System Resource*. Proc. of the 10th Intl. Conf. on Architectural Support for Programming Languages and Operating Systems, San Jose, USA.

[21] S. Azodolmolk et al. An *analytical model for software defined networking: A network calculus-based approach*. Proc. of the Global Communications Conference, 2013.

[22] J. Son, A. V. Dastjerdi, R. N. Calheiros and R. Buyya. *SLA-Aware and Energy-Efficient Dynamic Overbooking in SDN-Based Cloud Data Centers*. IEEE Transactions on Sustainable Computing, 2(2):76-89, 2017.

[23] G. Bolch et al. *Queueing networks and Markov chains: modeling and performance evaluation with computer science applications*. John Wiley & Sons, 2006.

[24] M. Marschall. *Chef Infrastructure Automation Cookbook*. Packt Publishing Ltd, 2013.

[25] Distributed Management Task Force. *Open Virtualization Format*. White Paper, June 2009. URL: www.dmtf.org/standards/published_documents/DSP2017_1.0.0.pdf

[26] Apache Giraph. http://giraph.apache.org/

[27] R. Chen et al. *Improving large graph processing on partitioned graphs in the Cloud*. Proceedings of the Third ACM Symposium on Cloud Computing, ACM, USA, 2012.

[28] D. Hardt. *The OAuth 2.0 Authorization Framework*. Internet Engineering Task Force (IETF) RFC 6749, 2012.

[29] R. Jain and S. Paul, *Network virtualization and software defined networking for cloud computing: a survey*, IEEE Communications Magazine, 51(11): 24-31, November 2013.

[30] J. Son, A. V. Dastjerdi, R. N. Calheiros, X. Ji, Y. Yoon and R. Buyya. *CloudSimSDN: Modeling and Simulation of Software-Defined Cloud Data Centers.* Proc. of the 15th IEEE/ACM International Symposium on Cluster, Cloud and Grid Computing (CCGrid 2015), Shenzhen, China, 2015.

[31] R. N. Calheiros, R. Ranjan, A. Beloglazov, C. A. De Rose and R. Buyya. *CloudSim: a toolkit for modeling and simulation of cloud computing environments and evaluation of resource provisioning algorithms,* Software: Practice and Experience, Wiley, 2011, 41, 23-50

[32] J. Son. *Integrated provisioning of compute and network resources in Software-Defined Cloud Data Centers*. PhD Thesis, The University of Melbourne, 2018.

[33] J. Son and R. Buyya, *A Taxonomy of Software-Defined Networking (SDN)-Enabled Cloud Computing*, ACM Computing Surveys, 51(3):1-36, ACM Press, USA, May 2018.